\documentclass{article}

     \usepackage[final]{neurips_2019_ml4h}

\usepackage[utf8]{inputenc} 
\usepackage[T1]{fontenc}    
\usepackage{hyperref}       
\usepackage{url}            
\usepackage{booktabs}       
\usepackage{amsfonts}       
\usepackage{nicefrac}       
\usepackage{microtype}      

\usepackage{wrapfig}

\usepackage{amsthm,amsmath}
\usepackage{amsbsy}
\usepackage{color}

\usepackage[all]{xy}

\usepackage{epsf}
\usepackage{epsfig}

\newcommand{\bfX}{\boldsymbol{X}}

\newcommand{\bfZ}{\boldsymbol{Z}}
\newcommand{\bfW}{\boldsymbol{W}}

\newcommand{\ci}{\perp\!\!\!\perp}
\newcommand{\nci}{\not\perp\!\!\!\perp}

\newtheorem{theorem}{Theorem}

\newcommand{\beginsupplement}{%
        \setcounter{table}{0}
        \renewcommand{\thetable}{S\arabic{table}}%
        \setcounter{figure}{0}
        \renewcommand{\thefigure}{S\arabic{figure}}%
     }

\title{Causality-based tests to detect the influence of confounders on mobile health diagnostic applications: a comparison with restricted permutations}

\author{%
  Elias Chaibub Neto$^*$ \\
  Meghasyam Tummalacherla, Lara Mangravite, Larsson Omberg \\
  \texttt{$^*$elias.chaibub.neto@sagebase.org} \\
  Sage Bionetworks \\
  Seattle, WA 98121 \\
}

\begin{document}

\maketitle

\begin{abstract}
Machine learning applications are frequently plagued with confounders that can impact the generalizability of the learners. In clinical settings, demographic variables often play the role of confounders. Confounding is especially challenging in remote digital health studies where the participants self-select to enroll in the study, thereby making it hard to balance the demographic characteristics of participants.

One effective method to combat confounding is to match samples with respect to the confounding variables in order to improve the balance of the data. This approach, however, leads to smaller datasets and hence negatively impact the inferences drawn from the learners. Alternatively, confounding adjustment methods that make more efficient use of the data (such as inverse probability weighting) usually rely on modeling assumptions, and it is unclear how robust these methods are to violations of these assumptions. This realization has motivated the development of restricted permutation (Good 2000) approaches to quantify the influence of observed confounders on the predictive performance of machine learning models and evaluate whether confounding adjustment methods are working as expected (Chaibub Neto et al 2018, 2019a).

Here, we build over previous work by Chaibub Neto et al (2018, 2019), in two important ways. First, we show that the use of restricted permutations to estimate the contribution of a confounder to the predictive performance of a learner can generate biased results. (In particular, we show that the restricted permutation approach can only correctly quantify the influence of confounders when the total association between disease labels and features is due to the confounders alone.)  Second, as an alternative to the use of restricted permutations, we propose a simple causality based approach to detect the influence of confounders on the predictive performance of diagnostic classifiers. We illustrate the application of our causality-based approach to data collected from mHealth study in Parkinson's disease.
\end{abstract}

\section{Restricted permutations from a causal perspective}

Here, we describe the restricted permutation approach (Chaibub Neto et al 2018, 2019a) from a causal perspective and show that its use to quantify the influence of confounders on the predictive performance of classifiers can generate biased results. (See Section 5.1 in the Appendix for a brief background on the causality definitions used in this paper.) For simplicity, we focus on the linear case where Wright's path coefficients (Wright 1934) directly provide a connection between causal effects and statistical associations.

As shown by Chaibub Neto et al (2018, 2019a), restricted permutations (Good 2000) can be used to detect the influence of confounders on the predictive performance of a learner. The key idea behind the approach is to shuffle the response data within the levels of a categorical/ordinal confounder. For instance, for a categorical confounder such as gender, a restricted permutation of the label data is obtained by shuffling the labels of the female subjects among themselves, and separately shuffling the labels of the males among themselves. Observe that the association between the confounder and the labels is perfectly preserved by the restricted permutation process. Furthermore, because the restricted permutations only shuffle the label data, we have that the associations between the confounder and the features are clearly preserved, as well. Only the associations between the shuffled labels and the features are modified by the restricted permutations (since the restricted permutations only preserve the association between the features and label data that is mediated by the confounder).

Next, we show the formal connection between restricted permutations, covariances/partial covariances, and Wright's path coefficients. The relationship between covariances, partial covariances, and restricted permutations is established by Theorem 1. (See Section 5.2 in the Appendix for the proof.)
\begin{theorem}
Let $Y^\ast$ represent a restricted permutation of the discrete labels $Y$ with respect to a discrete confounder variable $A$. Then,
\begin{equation}
E_{\pi^\ast}\left[ Cov(X, Y^\ast) \right] \, = \, Cov(X, Y) \, - \, Cov(X, Y \mid A) \, = \, Cov(X, A) \, Cov(Y, A)/Var(A)~.
\label{eq:restrpermcovpcovrelation2}
\end{equation}
where $E_{\pi^\ast}$ represents the expectation with respect to the restricted permutation null distribution, and $Cov(X, Y \mid A)$ represents the partial covariance of $X$ and $Y$ given $A$.
\end{theorem}

Theorem 1 shows that the expectation of the restricted permutation null distribution of the covariance operator can be computed analytically, and corresponds to the difference between the marginal and partial covariances of $X$ and $Y$, or, equivalently, to $Cov(X, A)Cov(Y, A)/Var(A)$.

The connection between covariances and causal effects, on the other hand, follows directly from Wright's method of path analysis. For instance, the application of Wright's path analysis to the causal diagram underlying sensor-based mobile health studies (presented in Figure \ref{fig:examplesPC1}, and fully described in Section 5.1.1 of the Appendix), combined with the result of Theorem 1 shows that,
\begin{align}
E_{\pi^\ast}[Cov(X, Y^\ast)] &= Cov(X, A) Cov(Y, A)/Var(A) = Cov(X, A) Cov(Y, A) \nonumber \\
&= (\theta_{XA} + \theta_{XY} \, \theta_{AY}) \, \theta_{AY} = \theta_{XA} \, \theta_{AY} + \theta_{XY} \, \theta_{AY}^2~.
\end{align}
The above result clearly shows that the average of the restricted permutation null distribution, $E_{\pi^\ast}[Cov(X, Y^\ast)]$, provides a biased estimate of the association that is contributed by the confounder alone (since $E_{\pi^\ast}[Cov(X, Y^\ast)]$ is given by $\theta_{XA} \theta_{AY} + \theta_{XY} \theta_{AY}^2$, rather than by $\theta_{XA} \, \theta_{YA}).$\footnote{Note that for the causal graph in Figure \ref{fig:examplesPC1} we have that the total association between $X$ and $Y$, namely, $Cov(X, Y) = \theta_{XY} + \theta_{XA} \theta_{YA}$, can be decomposed into the association generated by the direct causal effect of $Y$ on $X$ (i.e., $\theta_{XY}$), and the indirect/spurious association generated by $A$ (i.e., $\theta_{XA} \theta_{YA}$).}
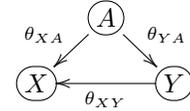
\begin{wrapfigure}{r}{0.2\textwidth}
\vskip -0.1in
$$
\xymatrix@-1.0pc{
 & *+[F-:<10pt>]{A} \ar[dl]_{\theta_{XA}} \ar[dr]^{\theta_{YA}} &  \\
*+[F-:<10pt>]{X} && *+[F-:<10pt>]{Y} \ar[ll]^{\theta_{XY}} \\
}
$$
\vskip -0.1in
  \caption{Causal diagram for sensor based mobile health studies.}
  \label{fig:examplesPC1}
  \vskip -0.1in
\end{wrapfigure}
Only in situations where $\theta_{XY} = 0$ we have that $E_{\pi^\ast}[Cov(X, Y^\ast)]$ will provide an unbiased estimate of the association between $X$ and $Y$ that is contributed by the confounder $A$. (Of course, we also obtain an unbiased estimate in situations where $\theta_{YA} = 0$, but this represents the uninteresting case where the classifier is not confounded.) Section 5.3 in the Appendix illustrates these points based on synthetic data generated from the causal model in Figure \ref{fig:examplesPC1}.

While these results illustrate the problem using the covariance metric and in the special case of linear models, it is clear that the same issue will also affect any other association metrics in linear or non-linear settings. Furthermore, in the context of machine learning applications, these illustrations also show that the location of the restricted permutation null should not generally be used to estimate the contribution of confounders to the predictive performance of a learner, as proposed by Chaibub Neto et al (2018, 2019a). [Restricted permutations can, nonetheless, still be safely used in prediction problems that are genuinely associational in nature such as computer vision tasks (Chaibub Neto 2019b).]

\section{Causality based tests to detect confounding in the predictive performance of ML algorithms}

In this section, we present an alternative approach to evaluate if the predictive performance of a classifier is being influenced by confounders. The key idea is to represent the classification task as a causal graph, and compare the conditional independence relations predicted by the application of the d-separation criterion to the graph against the conditional independence relations observed in the data. But, before we introduce the test, we first need to describe how the prediction scores, $\hat{R}_{ts}$, generated by a probabilistic classifier can be interpreted as the output of structural causal models (Pearl, 2000).

\subsection{Causal diagram representation of the classification task in diagnostic mobile health}

\begin{wrapfigure}{r}{0.55\textwidth}
\vskip -0.1in
{\scriptsize
$$
\xymatrix@-1.2pc{
*+[F-:<10pt>]{\bfX_{tr}} \ar@[red][dr] & *+[F-:<10pt>]{Y_{tr}} \ar@[red][d] & *+[F-:<10pt>]{\bfX_{ts}} \ar@[red][d] & *+[F-:<10pt>]{A_{tr}} \ar[d] \ar@/^1pc/@{<->}[r] & *+[F-:<10pt>]{Y_{tr}} \ar@[red][d] \ar[dl] &  & *+[F-:<10pt>]{A_{ts}} \ar[d] & *+[F-:<10pt>]{Y_{ts}} \ar[dl] \ar@/_1pc/@{<->}[l] \\
(a) & *+[F-:<10pt>]{M} \ar@[red][r] & *+[F-:<10pt>]{\hat{R}_{ts}} & *+[F-:<10pt>]{\bfX_{tr}} \ar@[red][r] & *+[F-:<10pt>]{M} \ar@[red][r] & *+[F-:<10pt>]{\hat{R}_{ts}} & *+[F-:<10pt>]{\bfX_{ts}} \ar@[red][l] & (b) \\
}
$$}
\vskip -0.1in
  \caption{Causal diagram of a mobile health classification task.}
  \label{fig:classdags}
  \vskip -0.1in
\end{wrapfigure}
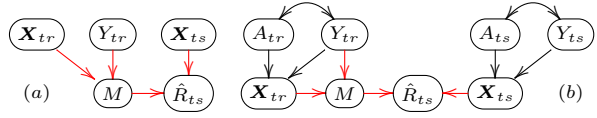

Training a classifier usually corresponds to creating a model, $M$, which corresponds to a (possibly stochastic) function, $g_1$, of the training data, $M = g_1(Y_{tr}, \bfX_{tr})$. (The boldface notation for $\bfX_{tr}$ represents the fact that we might have multiple features.) The predicted probability scores (i.e., the probabilities that each test example belongs to the positive class, $\hat{R}_{ts}$) are generated by applying model $M$ to the test set features. Usually, $\hat{R}_{ts}$ corresponds to a deterministic function $\hat{R}_{ts} = g_2(M, \bfX_{ts})$. (For instance, in the logistic regression model, $g_2$ corresponds to the logistic function.) Figure \ref{fig:classdags}a presents the causal diagram representation of the data generation process giving rise to the predictions. Figure \ref{fig:classdags}b presents the full causal diagram of the data generation process (black arrows) together with the generation process for the predictions (red arrows). Observe that because the confounder $A$ is not included as a feature in the classifier, the trained model is only a function of the training features, $\bfX_{tr}$, and the training labels, $Y_{tr}$.

\subsection{Statistical tests to detect confounding in ML predictions}

Now, we describe how to leverage the conditional independence relations spanned by the causal graphs associated with different confounding adjustment approaches.
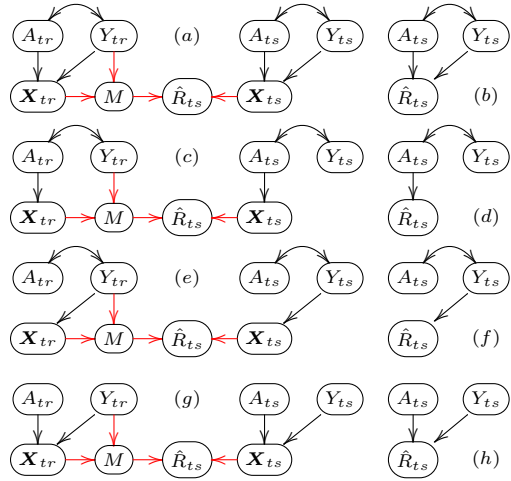
\begin{wrapfigure}{r}{0.475\textwidth}
\vskip -0.05in
{\scriptsize
$$
\xymatrix@-1.2pc{
*+[F-:<10pt>]{A_{tr}} \ar[d] \ar@/^1pc/@{<->}[r] & *+[F-:<10pt>]{Y_{tr}} \ar@[red][d] \ar[dl] & (a) & *+[F-:<10pt>]{A_{ts}} \ar[d] & *+[F-:<10pt>]{Y_{ts}} \ar[dl] \ar@/_1pc/@{<->}[l] & *+[F-:<10pt>]{A_{ts}} \ar[d] & *+[F-:<10pt>]{Y_{ts}} \ar[dl] \ar@/_1pc/@{<->}[l]  \\
*+[F-:<10pt>]{\bfX_{tr}} \ar@[red][r] & *+[F-:<10pt>]{M} \ar@[red][r] & *+[F-:<10pt>]{\hat{R}_{ts}} & *+[F-:<10pt>]{\bfX_{ts}} \ar@[red][l] & & *+[F-:<10pt>]{\hat{R}_{ts}} & (b) \\
*+[F-:<10pt>]{A_{tr}} \ar[d] \ar@/^1pc/@{<->}[r] & *+[F-:<10pt>]{Y_{tr}} \ar@[red][d] & (c) & *+[F-:<10pt>]{A_{ts}} \ar[d] & *+[F-:<10pt>]{Y_{ts}} \ar@/_1pc/@{<->}[l] & *+[F-:<10pt>]{A_{ts}} \ar[d] & *+[F-:<10pt>]{Y_{ts}} \ar@/_1pc/@{<->}[l]  \\
*+[F-:<10pt>]{\bfX_{tr}} \ar@[red][r] & *+[F-:<10pt>]{M} \ar@[red][r] & *+[F-:<10pt>]{\hat{R}_{ts}} & *+[F-:<10pt>]{\bfX_{ts}} \ar@[red][l] & & *+[F-:<10pt>]{\hat{R}_{ts}} & (d) \\
*+[F-:<10pt>]{A_{tr}} \ar@/^1pc/@{<->}[r] & *+[F-:<10pt>]{Y_{tr}} \ar@[red][d] \ar[dl] & (e) & *+[F-:<10pt>]{A_{ts}} & *+[F-:<10pt>]{Y_{ts}} \ar[dl] \ar@/_1pc/@{<->}[l] & *+[F-:<10pt>]{A_{ts}} & *+[F-:<10pt>]{Y_{ts}} \ar[dl] \ar@/_1pc/@{<->}[l]  \\
*+[F-:<10pt>]{\bfX_{tr}} \ar@[red][r] & *+[F-:<10pt>]{M} \ar@[red][r] & *+[F-:<10pt>]{\hat{R}_{ts}} & *+[F-:<10pt>]{\bfX_{ts}} \ar@[red][l] & & *+[F-:<10pt>]{\hat{R}_{ts}} & (f) \\
*+[F-:<10pt>]{A_{tr}} \ar[d] & *+[F-:<10pt>]{Y_{tr}} \ar@[red][d] \ar[dl] & (g) & *+[F-:<10pt>]{A_{ts}} \ar[d] & *+[F-:<10pt>]{Y_{ts}} \ar[dl] & *+[F-:<10pt>]{A_{ts}} \ar[d] & *+[F-:<10pt>]{Y_{ts}} \ar[dl]  \\
*+[F-:<10pt>]{\bfX_{tr}} \ar@[red][r] & *+[F-:<10pt>]{M} \ar@[red][r] & *+[F-:<10pt>]{\hat{R}_{ts}} & *+[F-:<10pt>]{\bfX_{ts}} \ar@[red][l] & & *+[F-:<10pt>]{\hat{R}_{ts}} & (h) \\
}
$$}
\vskip -0.1in
  \caption{Causal diagram representations of confounded and unconfounded classification tasks.}
  \label{fig:classificationcausaldiagram}
  \vskip -0.2in
\end{wrapfigure}
This can be used to evaluate if the adjustment was able to prevent the classifier from learning the confounding signal.

Figure \ref{fig:classificationcausaldiagram} presents the causal diagrams of classification tasks in situations where $A$ is a confounder (panels a/b and c/d), as well as, when it is not (panels e/f and g/h). Panel a presents the full causal diagram in the case where the causal relation $Y \rightarrow \bfX$ is confounded by $A$. Figure \ref{fig:classificationcausaldiagram}b presents a simplified diagram focusing only on the prediction scores, test set labels and confounders, where the causal paths $A_{ts} \rightarrow \bfX_{ts} \rightarrow \hat{R}_{ts}$ and $A_{ts} \leftrightarrow Y_{ts} \rightarrow \bfX_{ts} \rightarrow \hat{R}_{ts}$ are replaced by their collapsed versions $A_{ts} \rightarrow \hat{R}_{ts}$, $A_{ts} \leftrightarrow Y_{ts} \rightarrow \hat{R}_{ts}$, respectively. (Panels c and d show the causal diagram in the case where the association between the labels and the features is due only to the confounders.)

Panels e and f, on the other hand, show the causal diagram associated with confounding adjustment methods that aim to remove the association between the features and the confounders (e.g., residualization of features), while panels g and h show the causal diagram associated with confounding adjustment methods that aim to remove the association between confounders and the disease labels (e.g., matching or inverse probability weighting). Of course, these causal diagrams encode the important assumption that the confounding adjustment has worked as expected.

In order to evaluate if this is indeed the case, we need to compare the conditional independence relations predicted by d-separation against the conditional independence relations observed in the data. Assuming faithfulness (Spirtes et al 2000, Pearl 2000) of the probability distribution to causal diagram, we have that the following conditional (in)dependencies relations hold on panels b, d, f, and h of Figure \ref{fig:classificationcausaldiagram} (according to the d-separation criterion),
\begin{align}
\mbox{panel b:} \hspace{0.3cm} &{\hat{R}_{ts}} \nci Y_{ts}~, \; {\hat{R}_{ts}} \nci A_{ts}~, \; A_{ts} \nci Y_{ts}~, \; {\hat{R}_{ts}} \nci Y_{ts} \mid A_{ts}~, \; {\hat{R}_{ts}} \nci A_{ts} \mid Y_{ts}~, \label{eq:pred.CIs.confounding} \\
\mbox{panel d:} \hspace{0.3cm} &{\hat{R}_{ts}} \nci Y_{ts}~, \; {\hat{R}_{ts}} \nci A_{ts}~, \; A_{ts} \nci Y_{ts}~, \; {\hat{R}_{ts}} \ci Y_{ts} \mid A_{ts}~, \; {\hat{R}_{ts}} \nci A_{ts} \mid Y_{ts}~, \\
\mbox{panel f:} \hspace{0.3cm} &{\hat{R}_{ts}} \nci Y_{ts}~, \; {\hat{R}_{ts}} \nci A_{ts}~, \; A_{ts} \nci Y_{ts}~, \; {\hat{R}_{ts}} \nci Y_{ts} \mid A_{ts}~, \; {\hat{R}_{ts}} \ci A_{ts} \mid Y_{ts}~, \label{eq:pred.CIs.counter} \\
\mbox{panel h:} \hspace{0.3cm} &{\hat{R}_{ts}} \nci Y_{ts}~, \; {\hat{R}_{ts}} \nci A_{ts}~, \; A_{ts} \ci Y_{ts}~, \; {\hat{R}_{ts}} \nci Y_{ts} \mid A_{ts}~, \; {\hat{R}_{ts}} \nci A_{ts} \mid Y_{ts}~. \label{eq:pred.CIs.match}
\end{align}
Hence, we can use conditional independence (CI) tests (comparing the null hypotheses of statistical independence against the alternative of statistical association) in order to evaluate the effectiveness of a confounding adjustment. For instance, if matching or inverse probability weighting is working as expected we have that the CI test $H_0: A_{ts} \ci Y_{ts}$ vs $H_1: A_{ts} \nci Y_{ts}$ should show that $A_{ts}$ is marginally independent of $Y_{ts}$, while all the other 4 CI tests evaluating the remaining CI relations in (\ref{eq:pred.CIs.match}) should detect statistical dependencies. (See Section 5.4 in the Appendix for practical considerations on the use of linear and non-linear CI tests.)

\section{Analysis of the mPower data}
\begin{wrapfigure}{r}{0.67\textwidth}
\vskip -0.1in
\includegraphics[width=\linewidth]{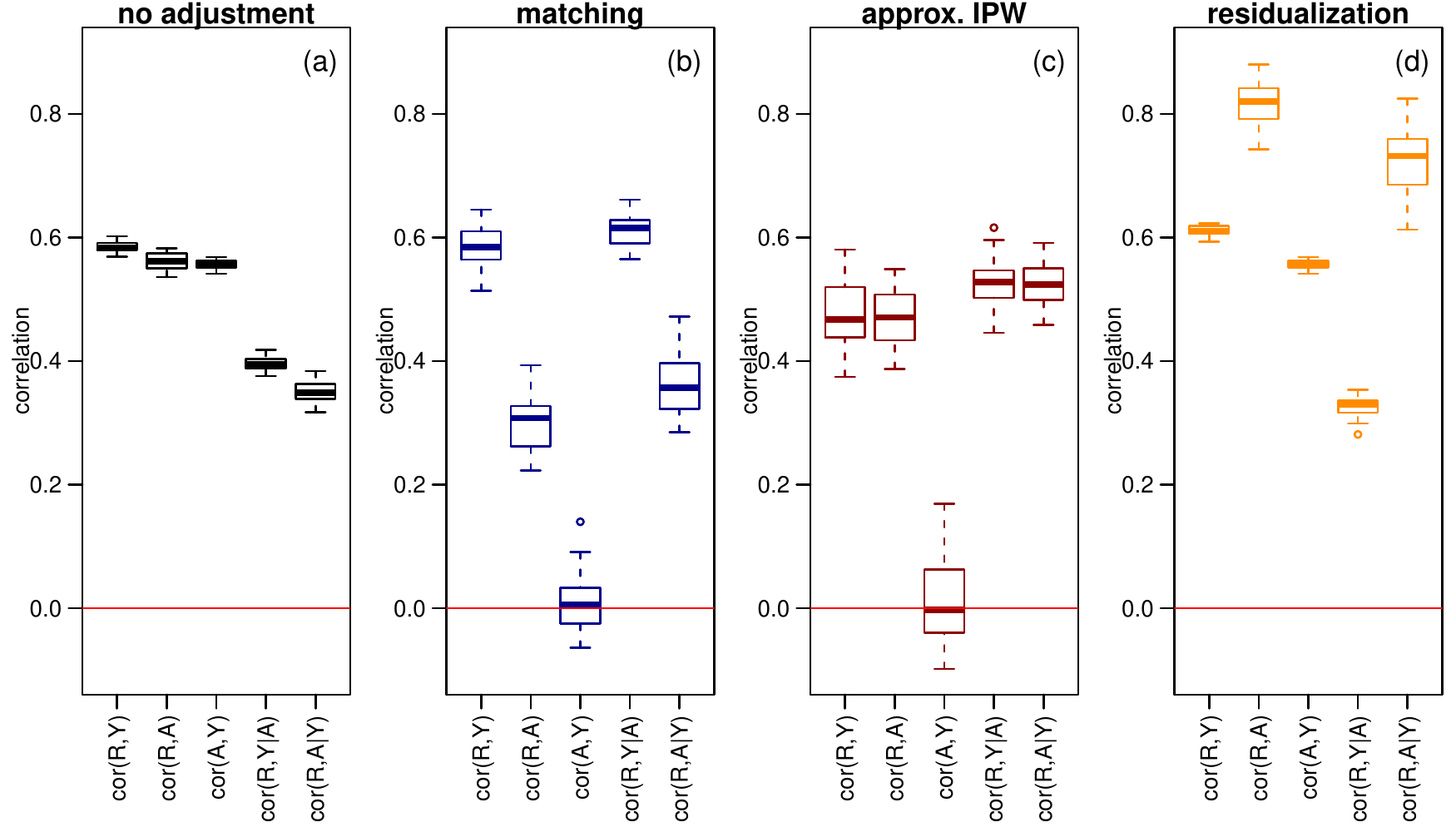}
\vskip -0.1in
\caption{Effectiveness of confounding adjustment for the ``no adjustment", matching, approximate IPW, and feature residualization methods in the mPower data.}
\label{fig:main.figure2}
\vskip -0.1in
\end{wrapfigure}

\vskip -0.1in
Here, we illustrate the application of our causality-based tests to three confounding adjustment methods, namely: sample matching, approximate inverse probability weighting (Linn et al 2016), and residualization of features. For the sake of comparison, we also report results for the original data without any adjustments. We reanalyzed the mPower data (Bot et al 2016) and investigated the influence of age on the predictive performance of random forest classifiers. (Section 5.5 of the Appendix provides further details on the data and modeling choices, as well as, a more detailed description of the results.) Figure \ref{fig:main.figure2} reports the observed marginal and partial Spearman correlation patterns which can be used to infer if the confounding adjustment methods are working as expected. In order to evaluate the statistical significance of each of these patterns we tested the null hypotheses,
\begin{equation}
{\hat{R}_{ts}} \ci Y_{ts}~, \;\;\; {\hat{R}_{ts}} \ci A_{ts}~, \;\;\; A_{ts} \ci Y_{ts}~, \;\;\; {\hat{R}_{ts}} \ci Y_{ts} \mid A_{ts}~, \;\;\; {\hat{R}_{ts}} \ci A_{ts} \mid Y_{ts}~,
\label{eq:null.hypotheses}
\end{equation}
and reported the p-values on Figure \ref{fig:main.fig.pvals}. Panel a reports the observed CI pattern for the original data (``no adjustment" method) and shows that, as expected, the observed pattern matches the pattern described in (\ref{eq:pred.CIs.confounding}). Panel b shows that the observed pattern for the matching adjustment exactly matches the CI pattern in (\ref{eq:pred.CIs.match}). Panel c reports the pattern for the approximate IPW adjustment, which again matches the CI pattern in (\ref{eq:pred.CIs.match}). However, statistical tests for non-linear associations did not support the CI relation $A_{ts} \ci Y_{ts}$, suggesting that the IPW method was not completely able to remove the age confounding (see Section 5.5 for further details). Finally, panel d reports the observed pattern for the residualization adjustment and shows that this method failed to prevent the classifier from learning the age signal. (Note that we would have expected to see the CI patterns in (\ref{eq:pred.CIs.counter}), had this adjustment worked as expected.)

\section{Discussion}
\vskip -0.1in
Confounding is a serious issue in mobile health, and the development of rigorous methods to evaluate confounding adjustment approaches has key practical importance. Here, we improve over previous work by Chaibub Neto et al (2018, 2019a) on the assessment of confounding in two important ways. First, we show that using restricted permutations to estimate the contribution of a confounder to the predictive performance of a learner can generate biased results. Second, we propose a simple causality-based approach to tackle this problem.

\textbf{Acknowledgements.} This work has been funded by the Robert Wood Johnson Foundation. The data was contributed by users of the Parkinson mPower mobile application as part of the mPower study developed by Sage Bionetworks and described in Synapse [doi:10.7303/syn4993293].

\textbf{Author contributions.} ECN: conceptualization, methodology, formal analysis, and data curation. MT: data curation. LM: funding acquisition. LO: project administration.

\clearpage

\beginsupplement

\section{Appendix}

\subsection{Preliminaries}

\subsubsection{Causal diagram representation of sensor-based mobile health studies}

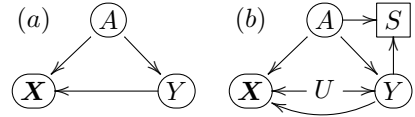
\begin{wrapfigure}{r}{0.4\textwidth}
$$
\xymatrix@-1.0pc{
(a) & *+[F-:<10pt>]{A} \ar[dl] \ar[dr] & & (b) & *+[F-:<10pt>]{A} \ar[dl] \ar[r] \ar[dr] & *+[F]{S} \\
*+[F-:<10pt>]{\bfX} && *+[F-:<10pt>]{Y} \ar[ll] & *+[F-:<10pt>]{\bfX}  & U \ar[l] \ar[r] & *+[F-:<10pt>]{Y} \ar@/^0.75pc/[ll] \ar[u] \\
}
$$
  \caption{Causal diagram representation of the data generation process in a sensor-based mobile health study.}
  \label{fig:examples1}
\end{wrapfigure}

Figure \ref{fig:examples1} presents the causal diagram representation of the data generating process in a sensor-based mobile health study. Here, $Y$, $A$, and $\bfX$ represent, respectively, the disease labels, the confounder, and sensor-based features. For concreteness, suppose that $Y$ represents Parkinson's disease (PD) status, $A$ represents age, and $\bfX$ represent a set of features extracted from the accelerometer sensor. Panel a shows a simplified causal diagram where the arrow from $Y$ to $\bfX$ represents the fact that the disease status has a causal effect on the walking patterns of the subjects (note that because people with PD tend to experience difficulty to walk, their acceleration patterns tend to be distinct from control/healthy subjects). The arrow from $A$ to $\bfX$ shows that age also influences walking patterns (as old subjects tend to move slower than younger ones). Finally, the arrow from $A$ to $Y$ indicates that age is a risk factor for PD. Panel b shows a slightly more complicated (and more realistic) representation of the data generation process. First, it assumes that the causal effect of $Y$ on $\bfX$ might be also confounded by other un-measured confounders represented by $U$. Second, it also assumes that the causal influence of $A$ on $Y$ might also be confounded by selection mechanisms (Hernan, Hernandez-Diaz, and Robins 2004).
For instance, let $S$ represent a binary selection variable indicating whether a potential participant has enrolled or not in the study (i.e., $S = 1$ if the person enrolled, and $S = 0$ otherwise). Note that age influences enrollment in the study ($A \rightarrow S$) because younger people tend to be more technology savvy than older people (and tech savvy people are more likely to enroll in mobile health studies). Observe as well that disease status influences enrollment ($Y \rightarrow S$), since subjects suffering from a disease are usually more motivated to enroll in a study than controls. The squared frame around $S$ in panel b indicates that the analysis is conditional on the participants that actually enrolled in the study (i.e, the analysis is conditional on $S = 1$). As mentioned above, selection biases are a common issue in mobile health studies.

\subsubsection{Notation and key definitions}

Following Pearl (2000), we adopt a mechanism-based approach to causation, where the statistical information encoded in the joint probability distribution of a set of variables is supplemented by a \textit{directed acyclic graph} (DAG) describing our qualitative assumptions about the causal relation between the variables. (Figure \ref{fig:examples1} presents a couple DAG examples, encoding causal assumptions in the context of mobile health studies.) In this framework, the joint probability distribution over a set of variables $\bfZ = \{Z_1, \ldots, Z_p\}$ factorizes according to the causal DAG structure, $P\big(Z_1, \ldots, Z_p \big) = \prod_{j} P\big(Z_j \mid pa(Z_j)\big)$, where each element, $P\big(Z_j \mid pa(Z_j)\big)$, represents an autonomous mechanism describing the relationship between variable $Z_j$ and its parents, $pa(Z_j)$. A non-parametric representation of these elements is given by the set of \textit{structural causal models} $Z_j = f_{Z_j}(pa(Z_j), U_{Z_j})$, where $f_{Z_j}$ represents a function of the parents of $Z_j$ and of a random disturbance term $U_{Z_j}$.

We define a \textit{path} as any unbroken, nonintersecting sequence of edges in a DAG, which may go along or against the direction of the arrows. We say that a path is \textit{d-separated} or \textit{blocked} (Pearl, 2000) by a set of nodes $\bfW$ if and only if: (i) the path contains a chain $Z_j \rightarrow Z_m \rightarrow Z_k$ or a fork $Z_i \leftarrow Z_m \rightarrow Z_k$ such that the middle node $Z_m$ is in $\bfW$; or (ii) the path contains an collider $Z_j \rightarrow Z_m \leftarrow Z_k$ such that the middle node $Z_m$ is not in $\bfW$ and no descendant of $Z_m$ is in $\bfW$. Otherwise, the path is \textit{d-connected} or \textit{open}.

We say that a joint probability distribution over a set of variables is \textit{faithful} (Spirtes et al 2000, Pearl 2000) to a DAG representing the causal relationships between these variables if no conditional independence relations, other than the ones implied by the d-separation criterion are present. Following standard practice, variables representing error terms are usually not shown in the diagram, while observed variables are represented by round frames, and conditioned-on variables are represented by square frames. A double arrow represents an association between two variables.

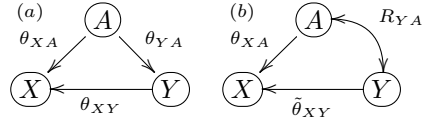
\begin{wrapfigure}{r}{0.4\textwidth}
$$
\xymatrix@-1.0pc{
^{(a)} & *+[F-:<10pt>]{A} \ar[dl]_{\theta_{XA}} \ar[dr]^{\theta_{YA}} & & ^{(b)} & *+[F-:<10pt>]{A} \ar[dl]_{\theta_{XA}} \ar@/^1pc/@{<->}[rd]^{R_{YA}} &  \\
*+[F-:<10pt>]{X} && *+[F-:<10pt>]{Y} \ar[ll]^{\theta_{XY}} & *+[F-:<10pt>]{X}  && *+[F-:<10pt>]{Y} \ar[ll]^{\tilde{\theta}_{XY}} \\
}
$$
  \caption{Path coefficients for the causal diagram in Figure \ref{fig:examples1}.}
  \label{fig:examplesPC}
\end{wrapfigure}

For linear structural equations, it is well known that the covariance (correlation) between any two standardized variables\footnote{Note that the linear model $Z^o_k = \mu_k + \Sigma_j \beta_{kj} Z^o_j + U^o_k$ can be be reparameterized into its equivalent standardized form $Z_k = \sum_{j} \theta_{kj} Z_j + U_k$, where $Z_k = (Z^o_k - E(Z^o_k))/\sqrt{Var(Z^o_k)}$ represent standardized variables with $E(Z_k) = 0$ and $Var(Z_k) = 1$; $\theta_{kj} = \beta_{kj} \sqrt{Var(Z^o_j)/Var(Z^o_k)}$ represent the path coefficients; and $U_k = U^o_k/\sqrt{Var(Z^o_k)}$ represent the standardized error terms. Note that for the standardized model, we have that the covariances are identical to the correlations - i.e., $Cov(Z_k, Z_j) = Cor(Z_k, Z_j)$ - since all variables have variance equal to 1.} in a causal diagram can be decomposed into a sum of products of path coefficients across all open paths connecting the two variables (Wright 1934). The path coefficient between two variables represents the causal effect of the parent variable on the child one. Figure \ref{fig:examplesPC} shows the path coefficients associated with the causal diagrams in Figure \ref{fig:examples1} (in the situation where $Y$ is continuous).

In Figure \ref{fig:examplesPC}a, $X$ and $Y$ are connected by the paths $Y \rightarrow X$ and $X \leftarrow A \rightarrow Y$, and their covariance is decomposed as,
$$
Cov(X, Y) = \underbrace{\theta_{XY}}_{X \leftarrow Y} + \underbrace{\theta_{XA} \, \theta_{YA}}_{X \leftarrow A \rightarrow Y}~,
$$
where the path coefficient $\theta_{XY}$ quantifies the contribution of the direct path $Y \rightarrow X$, whereas the product of path coefficients $\theta_{XA} \, \theta_{YA}$ quantifies the contribution of the backdoor path $X \leftarrow A \rightarrow Y$. Similarly, $Cov(X, A)$ and $Cov(A, Y)$ can be decomposed as,
$$
Cov(X, A) = \underbrace{\theta_{XA}}_{X \leftarrow A} + \underbrace{\theta_{XY} \, \theta_{YA}}_{X \leftarrow Y \leftarrow A}~, \hspace{0.3cm} \mbox{and} \hspace{0.3cm} Cov(A, Y) = \underbrace{\theta_{YA}}_{Y \leftarrow A}~,
$$
respectively. Furthermore, path analysis can also account for unspecified associations between (discrete) exogenous variables (Wright 1934). For instance, in Figure \ref{fig:examplesPC}b, we can represent the total association between $A$ and $Y$ by the covariance (correlation) $R_{YA}$ (which quantifies the joint contribution of the causal association induced by the path $A \rightarrow Y$ and the spurious association induced by the selection bias generated by the open collider path $A \rightarrow \fbox{S} \leftarrow Y$). In this case, the covariance decompositions are given by,
\begin{equation}
Cov(X, Y) = \underbrace{\tilde{\theta}_{XY}}_{X \leftarrow Y} + \underbrace{\theta_{XA} \, R_{YA}}_{X \leftarrow A \leftrightarrow Y}~, \hspace{0.1cm} Cov(X, A) = \underbrace{\theta_{XA}}_{X \leftarrow A} + \underbrace{\tilde{\theta}_{XY} \, R_{YA}}_{X \leftarrow Y \leftrightarrow A}~, \hspace{0.1cm} Cov(A, Y) = \underbrace{R_{YA}}_{Y \leftrightarrow A}~,
\end{equation}
where $\tilde{\theta}_{XY}$ represents a confounded causal effect given by $\tilde{\theta}_{XY} = \theta_{XY} + \theta_{XU} \theta_{YU}$, with $\theta_{XY}$ representing the real causal effect of $Y$ on $X$, and $\theta_{XU}$ and $\theta_{YU}$ representing the causal effects of the unmeasured confounder $U$ on $X$ and $Y$, respectively.

\subsection{Proof of Theorem 1}

\begin{proof}
By definition,
\begin{equation}
Cov(X, Y \mid A) \, = \, Cov(X, Y) \, - \, Cov(X, A) \, Var(A)^{-1} \, Cov(Y, A)~.
\label{eq:pcovdef}
\end{equation}
Hence, $Cov(X, Y^\ast \mid A) \, = \, Cov(X, Y^\ast_{A}) \, - \, Cov(X, A) \, Var(A)^{-1} \, Cov(Y^\ast, A)$ and
\begin{equation}
E_{\pi^\ast}\left[Cov(X, Y^\ast \mid A)\right] \, = \, E_{\pi^\ast}\left[Cov(X, Y^\ast)\right] \, - \, E_{\pi^\ast}\left[Cov(X, A) \, Var(A)^{-1} \, Cov(Y^\ast, A)\right]~.
\label{eq:expectationpcov}
\end{equation}

Now, observe that because the restricted permutations shuffle the response data within each level of the protected variable, it follows that (on average) the covariance of $X$ and $Y^\ast$ will be zero within each level of $A$, that is, $E_{\pi^\ast}\left[Cov(X, Y^\ast \mid A = a)\right] \, = \, 0$. Now, since this is true for all levels of $A$, it follows that, $E_{\pi^\ast}\left[Cov(X, Y^\ast \mid A)\right] \, = \, 0$. Hence, from eq. (\ref{eq:expectationpcov}) we have that,
\begin{align}
E_{\pi^\ast}\left[Cov(X, Y^\ast)\right] \, &= \, E_{\pi^\ast}\left[Cov(X, A) \, Var(A)^{-1} \, Cov(Y^\ast, A)\right] \nonumber \\
&= \, E_{\pi^\ast}\left[Cov(X, A) \, Var(A)^{-1} \, Cov(Y, A)\right] \nonumber \\
&= \, Cov(X, A) \, Var(A)^{-1} \, Cov(Y, A)~, \label{eq:equality}
\end{align}
where the second equality follows from the fact that $Cov(Y^\ast, A) = Cov(Y, A)$. Therefore, it follows from (\ref{eq:pcovdef}) and (\ref{eq:equality}) that, $E_{\pi^\ast}\left[Cov(X, Y^\ast)\right] \, = \, Cov(X, Y) \, - \, Cov(X, Y \mid A)$.
\end{proof}

\subsection{Restricted permutations - synthetic data illustrations}

As described in the main text, for the causal diagram representation of a sensor-based mobile health study (Figure \ref{fig:examplesPC1}), and in the special case that the relationships between the variables are linear, we have that the connection between restricted permutations and causal effects is summarized by,
\begin{equation}
E_{\pi^\ast}[Cov(X, Y^\ast)] = \theta_{XA} \, \theta_{AY} + \theta_{XY} \, \theta_{AY}^2~.
\end{equation}

\begin{figure}[!h]
\includegraphics[width=\columnwidth]{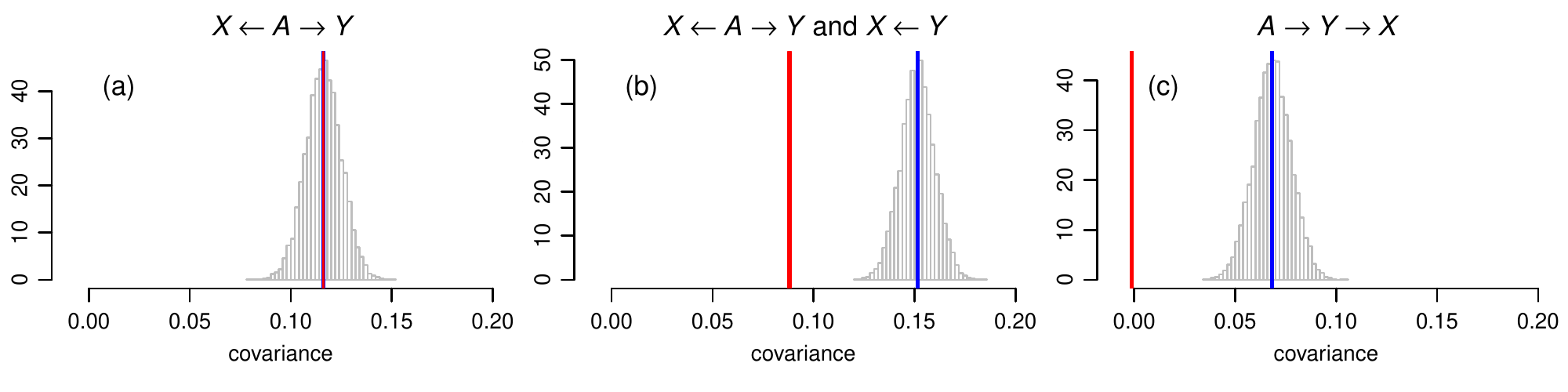}
\caption{Synthetic data examples where the restricted permutation null distribution correctly quantifies the influence of the confounder (panel a), as well as, when it does not (panels b and c).}
\label{fig:perm.nulls}
\end{figure}

The above equation clearly shows that the average of the restricted permutation null distribution, $E_{\pi^\ast}[Cov(X, Y^\ast)]$, provides a biased estimate of the association that is due to the confounder alone, i.e., $\theta_{XA} \, \theta_{YA}$. (Recall that the marginal covariance $Cov(X, Y) = \theta_{XY} + \theta_{XA} \theta_{YA}$ is decomposed into the association generated by the direct causal effect of $Y$ on $X$, $\theta_{XY}$, and the indirect/spurious association generated by $A$, $\theta_{XA} \theta_{YA}$.) Only in situations where $\theta_{XY} = 0$ we have that $E_{\pi^\ast}[Cov(X, Y^\ast)]$ will provide an unbiased estimate of the association between $X$ and $Y$ that is contributed by the confounder $A$. Figure \ref{fig:perm.nulls} illustrates this point based on synthetic data generated from the causal model in Figure \ref{fig:examplesPC1}. In all panels, the histogram represents the restricted permutation null, the blue line is set at $\hat{Cov}(X,A)\hat{Cov}(Y,A) = \hat{\theta}_{XA} \hat{\theta}_{AY} + \hat{\theta}_{XY} \hat{\theta}_{AY}^2$, and the red line is set at $\hat{\theta}_{XA} \hat{\theta}_{AY}$.

Figure \ref{fig:perm.nulls}a shows an example where we simulated data from the model in Figure \ref{fig:examplesPC1} using $\theta_{XY} = 0$. In this case, the red and blue lines match since $\hat{\theta}_{XY} \approx 0$, and the average of the restricted permutation null (blue line) correctly measures the spurious association due to the confounder alone, $\hat{\theta}_{XA} \hat{\theta}_{AY}$. Figures \ref{fig:perm.nulls}b and c, on the other hand, show two examples where this is not the case. In Figure \ref{fig:perm.nulls}b, we simulated data with all path coefficients different from zero, while in Figure \ref{fig:perm.nulls}c we simulated data with $\theta_{XA} = 0$. Note that in both cases the red and blue lines no longer match, showing that the location of the restricted permutation null does not correctly quantifies the contribution of the confounders to the association between $X$ and $Y$ (which is given by the red line).

For these synthetic data examples, we simulated data from the causal graph in Figure \ref{fig:examplesPC1} according to the model $A \sim Bernoulli(p)$, $Y = \beta_{YA} A + U_Y$, and $X = \beta_{XY} Y + \beta_{XA} A + U_X$, where $U_Y \sim N(0, \sigma^2_Y)$ and $U_X \sim N(0, \sigma^2_X)$. The data was then standardized so that the causal effects of $A$ on $Y$, $A$ on $X$ and $Y$ on $X$ where expressed in terms of the path coefficients, $\theta_{YA} = \beta_{YA}\sqrt{Var(A)/Var(Y)}$, $\theta_{XA} = \beta_{XA}\sqrt{Var(A)/Var(X)}$, and $\theta_{XY} = \beta_{XY}\sqrt{Var(Y)/Var(X)}$, where $Var(A) = p(1-p)$, $Var(Y) = \sigma^2_Y + \beta_{YA}^2 Var(A)$, $Var(X) = \sigma^2_X + \beta_{XA}^2 Var(A) + \beta_{XY}^2 Var(Y)$.

The path coefficients correspond to the regression coefficients of the standardized data and can be estimated directly from the linear model fit. For the first example in Figure \ref{fig:perm.nulls}a we simulated data with $\beta_{XA} = 0.75$, $\beta_{XY} = 0$, and $\beta_{YA} = 0.75$. For the example in Figure \ref{fig:perm.nulls}b we adopted $\beta_{XA} = \beta_{XY} = \beta_{YA} = 0.75$. For the third example in Figure \ref{fig:perm.nulls}c we adopted $\beta_{XA} = 0$, $\beta_{XY} = \beta_{YA} = 0.75$. In all examples we adopted $\sigma^2_X = \sigma^2_Y = 1$ and $p = 0.5$.

\subsection{Practical considerations for the conditional independence tests}

There is a rich literature on statistical tests for conditional independence (CI). In the particular case of linear models with Gaussian errors Pearson correlation and partial correlation can be used to test for marginal and conditional independencies. Partial rank correlations based on Spearman correlation offer a non-parametric alternative that can be used to relax the assumptions of Gaussian errors and strict linear associations (recall that Spearman correlation can pick-up monotone non-linear associations).

There has also been some progress towards the development of CI tests able to handle arbitrary non-linear associations and data distributions. These methods include kernel-based approaches (Fukumizu et al 2008, Zhang et al 2011) and approaches based on the explicit estimation of conditional densities (Su and White 2007, 2008). These methods, however, require continuous variables and cannot be directly applied to our problem (which also involves CI testing between continuous and binary variables). An alternative approach based on distance correlation (Szekely, Rizzo, Bakirov 2007) and partial distance correlation (Szekely and Rizzo 2014) can nonetheless be used to handle binary variables. The computation of (partial) distance correlations, however, can be slow when the sample size is large, and statistical tests are based on a permutation approach (which can be computationally intensive). Furthermore, while a zero distance correlation is equivalent to statistical independence (Szekely, Rizzo, Bakirov 2007), a zero partial distance correlation is not strictly equivalent to conditional independence (although, in practice, testing for zero partial distance correlation is often a good approximation for testing for conditional independence) (Szekely and Rizzo 2014).

The bottom line is that testing for CI is a non trivial task in non-linear settings, and no silver bullet test has yet been developed. Hence, in our illustrations we adopt (partial) rank correlation tests to perform a first pass of CI tests, and only use (partial) distance correlation to confirm if any conditional independencies detected by the (partial) rank correlation tests really hold in the data. Note that if the (partial) rank correlation test already rejected the null hypothesis of (conditional) independence, there is no need to calculate the computationally more expensive (partial) distance correlation test. However, whenever the (partial) rank correlation test fails to reject the null, then it is possible that non-monotone non-linear associations are still present, and it is necessary to employ the (partial) distance correlation test.

Finally, we point out that while the use of CI tests has a long history in causal discovery algorithms, such as the PC and FCI algorithms (Spirtes et al 2000), here we employ the CI tests in a different way. Namely, rather than trying to learn the causal graph from the data, we assume that the causal graph is already known and use CI tests to confirm if the CI relations predicted by the application of d-separation to the assumed causal graph are really holding in the data.

\subsection{Further details on the analysis of the mPower data}

As briefly described in the main text, we illustrate the application of our causality-based tests using three confounding adjustment methods, namely: sample matching, approximate inverse probability weighting (Linn et al 2016) based on the propensity score (Rosenbaum and Rubin 1983), and residualization of features. (For completeness, we also report results for the original data without any adjustments.) We reanalyzed the same Parkinson's disease mobile health dataset (Bot et al 2016) used by Chaibub Neto et al (2019a), and investigated the influence of age on the predictive performance of random forest classifiers.

The dataset consists of 50 features generated by a deep learning model trained on raw accelerometer data from a subset of the mPower data. (These features correspond to the winning submission of sub-challenge 1 of the Digital Biomarker Dream Challenge, and are publicly available at \texttt{https://www.synapse.org/\#!Synapse:syn10949406}). For these illustrations we focused on a subset of 3,944 subjects (796 cases and 3,148 controls) which where used to train the deep model. We trained random forests classifiers (with number of trees set to 500, and randomly sampling 7 features as candidates per node split) on 30 distinct random splits of the data and applied the adjustment methods to both training and test sets.

\begin{figure}[!h]
\includegraphics[width=\linewidth]{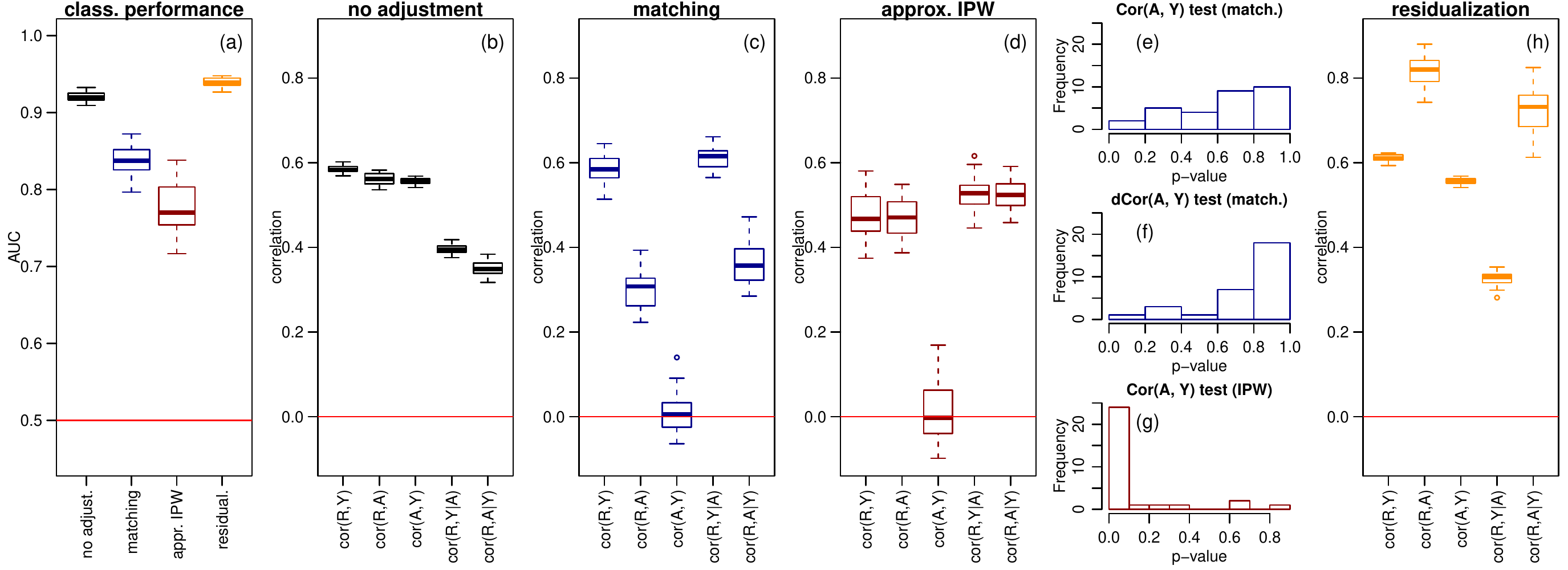}
\caption{Effectiveness of confounding adjustment for the ``no adjustment", matching, approximate IPW, and feature residualization methods in the mPower data.}
\label{fig:main.figure}
\end{figure}

\begin{figure}[!h]
\begin{center}
\includegraphics[width=3.6in]{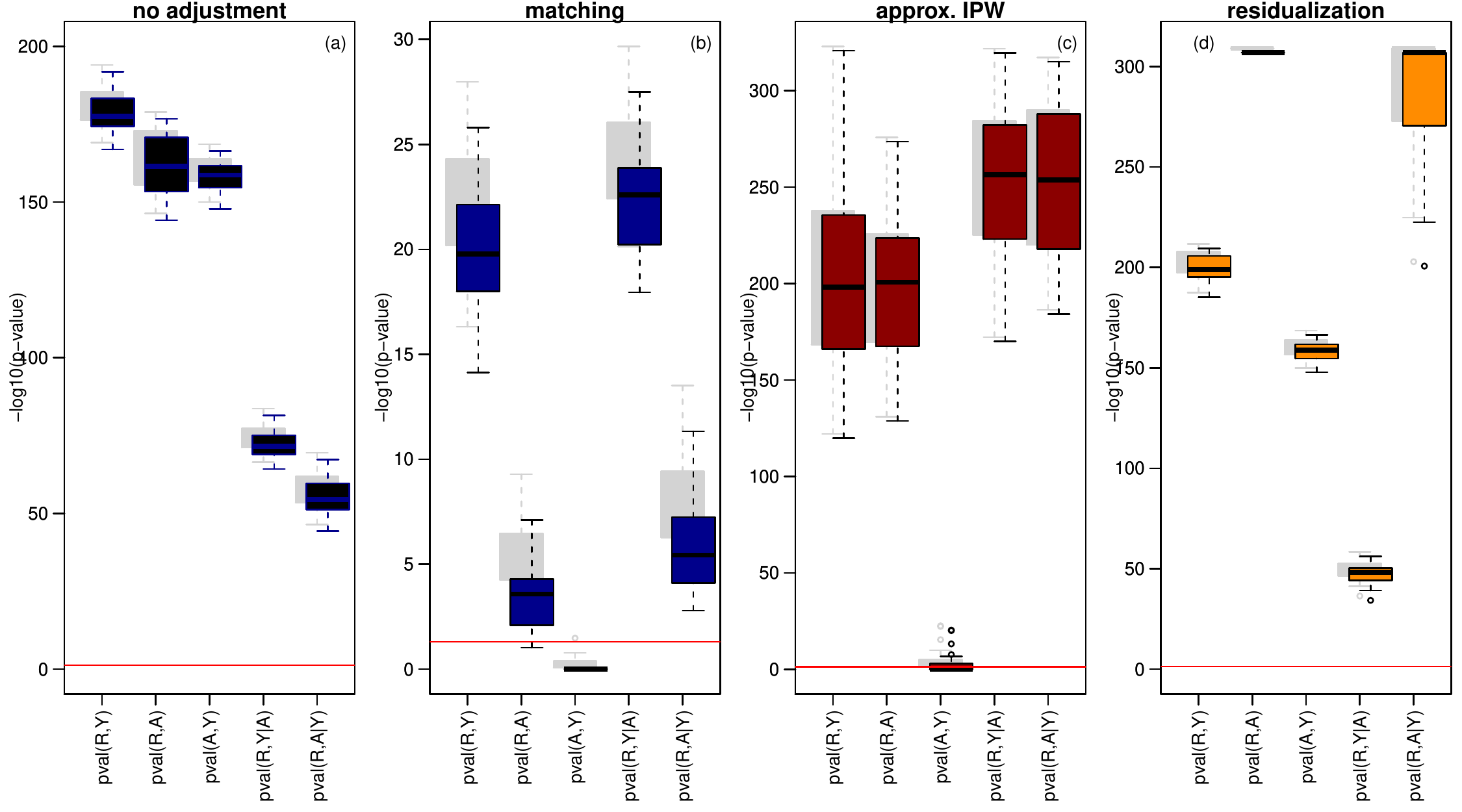}
\end{center}
\caption{P-values associated with the (partial) correlation tests (in -log10 scale). The red line is set at -log10(0.05). The grey boxplots show the raw p-values, while the colored boxplots show Bonferroni corrected p-values.}
\label{fig:main.fig.pvals}
\end{figure}

\begin{figure}[!h]
\begin{center}
\includegraphics[width=4.2in]{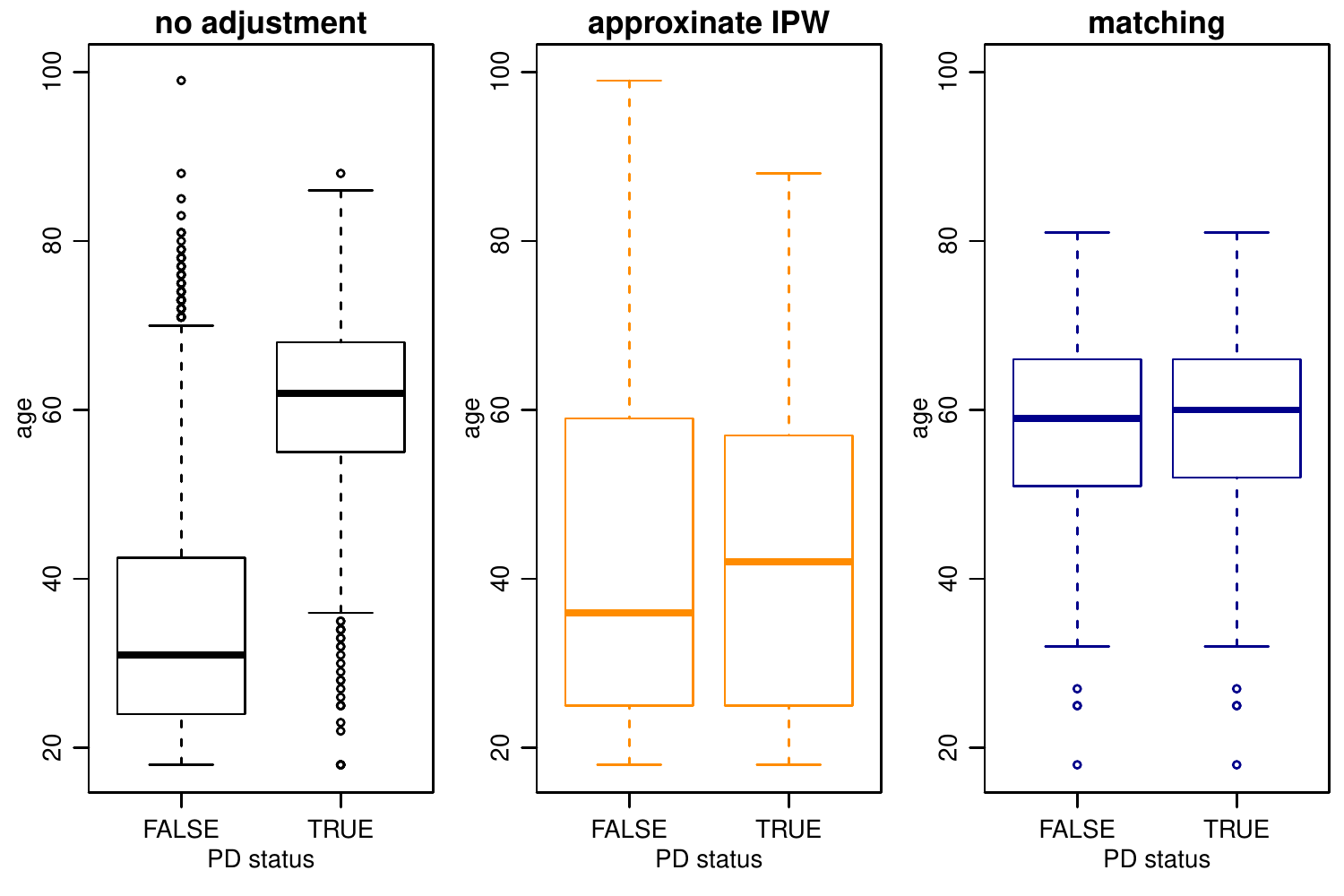}
\end{center}
\vskip -0.1in
\caption{Balance between age and PD status.}
\label{fig:balance.figure}
\vskip 0.2in
\end{figure}

\begin{figure}[!h]
\includegraphics[width=\linewidth]{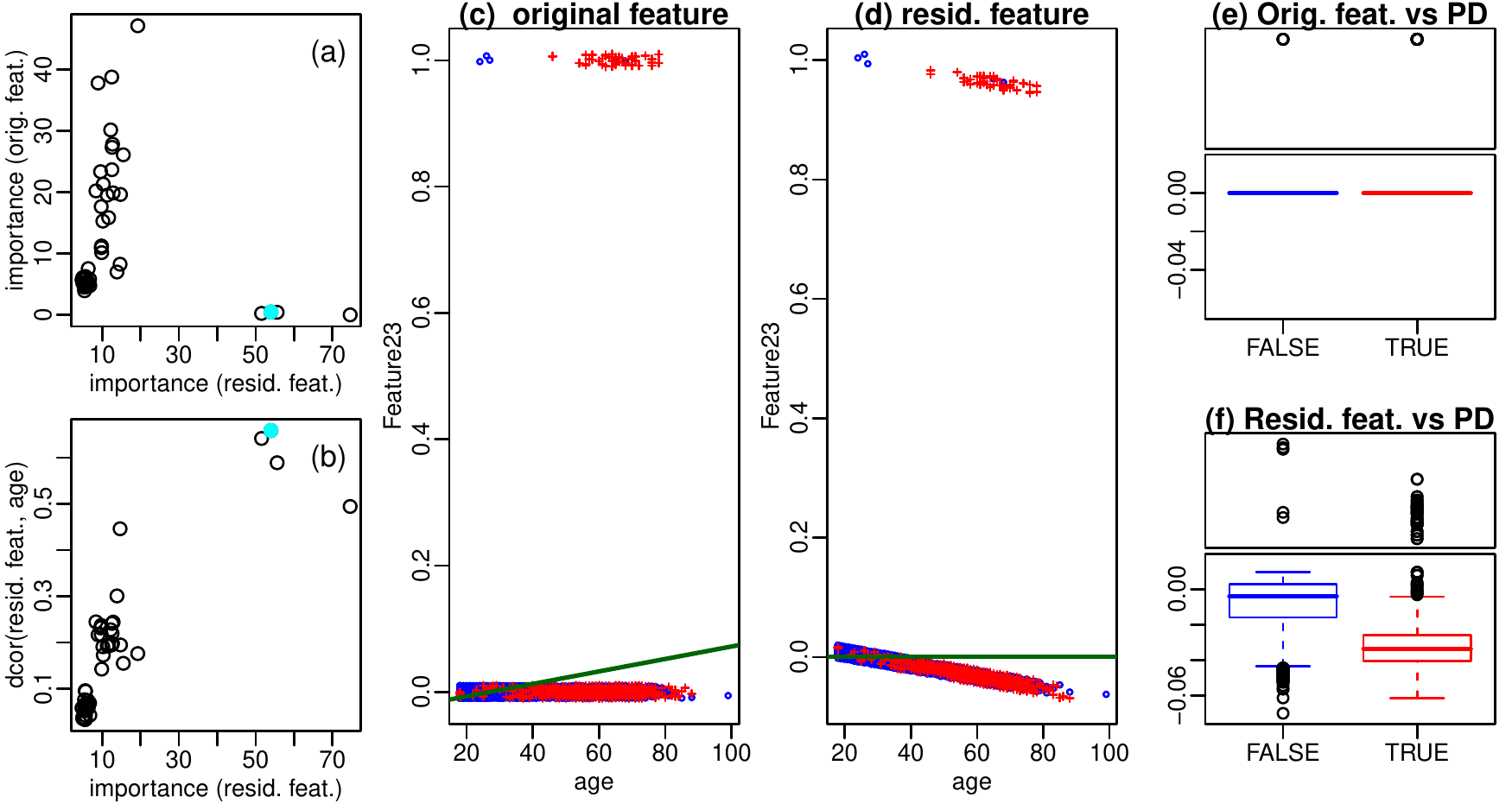}
\caption{Residualization artifact. Panel a shows a scatter plot of the importance metric of the random forest computed using the residualized features (x-axis) versus the original features (y-axis). The four dots in the bottom right show that the four least important features for the random forest trained with original features became the four most important features after residualization. Panel b shows a scatter plot of the residualized features importance score against the distance correlation between the residualized features and age. The plot shows that these four features turn out to have the strongest non-linear association with age. The cyan dot highlights the feature with the highest distance correlation with age (Feature23). Panel c plots Feature23 before residualization against age. (The plot actually shows jittered values as this feature is binary.) The blue and red dots represent control and case subjects, respectively. The green line shows the regression of the feature on age. Panel d plots the residualized version of Feature23 against age. Note how the residualization increases the association between the feature and age and between the feature and the disease labels (panels e and f), even though the correlation between the residualized feature and age is zero (as shown by the horizontal green line in panel d).}
\label{fig:resid.issue}
\end{figure}

Figure \ref{fig:main.figure} presents the results. Panel a shows the AUC scores for the original data (``no adjustment") and confounding adjustment methods. Panels b, c, d, and h show the marginal and partial Spearman correlations used to test the null hypotheses,
\begin{equation}
{\hat{R}_{ts}} \ci Y_{ts}~, \;\;\; {\hat{R}_{ts}} \ci A_{ts}~, \;\;\; A_{ts} \ci Y_{ts}~, \;\;\; {\hat{R}_{ts}} \ci Y_{ts} \mid A_{ts}~, \;\;\; {\hat{R}_{ts}} \ci A_{ts} \mid Y_{ts}~.
\label{eq:null.hypotheses}
\end{equation}
Panel b shows the results for the original data. Not surprisingly, application of the CI tests (Figure \ref{fig:main.fig.pvals}a) rejected all the null hypotheses in (\ref{eq:null.hypotheses}), suggesting that the predictive performance is confounded by age. (Note that the tests recovered the conditional independence pattern in (\ref{eq:pred.CIs.confounding}).)  Panel c shows the (partial) correlation estimates for the matching adjustment. Application of the correlation CI tests (Figure \ref{fig:main.fig.pvals}b) rejected all null hypotheses in (\ref{eq:null.hypotheses}) except for $A_{ts} \ci Y_{ts}$, in accordance with the expected CI pattern for an un-confounded classification task shown in (\ref{eq:pred.CIs.match}). Because the Spearman correlation might miss non-linear associations we also confirmed this statistical independency using distance correlation tests. The p-values for the correlation and distance correlation tests are shown in panels e and f. Panel d presents the correlations for the approximate IPW adjustment method. Similarly to matching, the correlation CI tests (Figure \ref{fig:main.fig.pvals}c) clearly rejected all null hypotheses except for $A_{ts} \ci Y_{ts}$, where the CI tests accepted this null hypothesis for a few of the data splits. (The correlation test p-values are presented in panel g). Application of the distance correlation tests did not confirm, however, this statistical independency (the permutation p-values based on 1,000 permutations were equal to 0.001 in all data splits). This suggests that the approximate IPW method was not completely able to prevent the random forest classifier from learning the age signal. This observation is further corroborated by Figure \ref{fig:balance.figure} which shows that while the IPW method improved the balance in the data, it still did not manage to completely remove the association between age and disease status (as was done by the matching adjustment). Panel h shows the (partial) correlation estimates for the feature residualization method (where we regressed each feature on age and used the residuals as the features in the classifier). Application of the correlation CI tests (Figure \ref{fig:main.fig.pvals}d) rejected all the null hypotheses in (\ref{eq:null.hypotheses}), showing that the residualization failed to prevent the classifier from learning the age signal. (Note that we would have expected to see the CI patterns in \ref{eq:pred.CIs.counter}, had this adjustment worked.) Observe that while at first sight this adjustment might seem to be working really well (note that the AUC scores in panel a are even higher than for the ``no adjustment") Supplementary Figure \ref{fig:resid.issue} shows that this improved performance is in reality due to an artifact that actually improves the ability of the classifier to learn the age signal rather than prevent learning it. This example illustrates how our tests can avoid these artifacts.

\end{document}